\documentclass[11pt]{article}
\usepackage{moriond,epsfig}

\usepackage{amsmath}

\usepackage{amsfonts,amsmath,amssymb,bm}

\def\be{\begin{equation}}
\def\ee{\end{equation}}
\def\bea{\begin{eqnarray}}
\def\eea{\end{eqnarray}}

\newcommand{\ud}{\mathrm{d}}

\begin{document}
\vspace*{4cm}
\title{Atom interferometry and the Einstein equivalence principle}

\author{Peter WOLF$\,^1$, Luc BLANCHET$\,^2$, Christian
  J. BORD\'E$\,^{1,3}$, \\Serge REYNAUD$\,^4$, Christophe
  SALOMON$\,^5$, Clande COHEN-TANNOUDJI$\,^5$}

\address{\vspace{0.2cm}$^1$LNE-SYRTE,
    Observatoire de Paris, CNRS, UPMC, France \\ $^2$GRECO, Institut
    d'Astrophysique de Paris, CNRS, UPMC, France \\ $^3$Laboratoire de
    Physique des Lasers, Universit\'{e} Paris 13, CNRS, France \\
    $^4$Laboratoire Kastler Brossel, CNRS, ENS, UPMC, France \\
    $^5$Laboratoire Kastler Brossel et Coll\`ege de France, CNRS, ENS,
    UPMC, France}

  \maketitle\abstracts{The computation of the phase shift in a symmetric atom
    interferometer in the presence of a gravitational field is
    reviewed. The difference of action-phase integrals between the two
    paths of the interferometer is zero for any Lagrangian which is at
    most quadratic in position and velocity. We emphasize that in a
    large class of theories of gravity the atom interferometer permits
    a test of the weak version of the equivalence principle (or
    universality of free fall) by comparing the acceleration of atoms
    with that of ordinary bodies, but is insensitive to that aspect of
    the equivalence principle known as the gravitational redshift or
    universality of clock rates.}


\noindent
The Einstein equivalence principle is at the basis of our
understanding of modern theories of gravity\,\cite{Will}. The weak
version of the equivalence principle, also called universality of free
fall (UFF), has been verified with high precision using torsion
balances\,\cite{Dicke,Braginsky,Schlamminger} and the Lunar laser
ranging\,\cite{Williams}. Atom interferometry experiments have also
yielded an important test of the UFF at the level $7\times10^{-9}$, by
comparing the acceleration of atoms with that of ordinary
bodies\,\cite{Peters,Mueller,refsyrte}.

The gravitational redshift of universality of clock rates (UCR) is the
least tested aspect of the equivalence principle. It is currently
known with $10^{-4}$ accuracy\,\cite{Vessot} and should be tested at
the level $10^{-6}$ in future space experiments with
clocks\,\cite{Cacciapuoti}. In this contribution we report
arguments\,\cite{WolfNat,WolfCQG} showing that a recent
claim\,\cite{Mueller} that atom interferometry experiments have
actually already tested the UCR at the level $7\times10^{-9}$ (thereby
improving the validity of the redshift by several orders of magnitude,
even with respect to future space experiments\,\cite{Cacciapuoti}), is
fundamentally incorrect. More recently, our arguments have received
support from several independent analyses\,\cite{Samuel,Giulini}.

In conventional clock experiments\,\cite{Vessot,Cacciapuoti}, the
measurement of the gravitational redshift uses two clocks $A$ and $B$
located at different heights in a gravitational field, and operating
at the frequency $\omega$ of an atomic transition. The two measured
frequencies $\omega_A$ and $\omega_B$ are continuously compared
through the exchange of electromagnetic signals. These measurements
rely on atomic spectroscopy (C{\ae}sium clocks, hydrogen masers,
optical clocks, etc.) or nuclear spectroscopy (like in the Pound-Rebka
experiment\,\cite{Pound}). The two clocks are put in devices
(experimental set-ups, rockets, satellites, etc.) that are classical
and whose trajectories can be measured by radio or laser ranging. The
atomic transition of $A$ and $B$ used as a frequency standard is
described \textit{quantum mechanically} but the motion of $A$ and $B$
in space can be described \textit{classically}. The motion of the two
clocks can thus be precisely measured and the contribution of the
special relativistic term (i.e. the Doppler effect) can be evaluated
and subtracted from the total frequency shift to get a test of the
gravitational redshift.

In the proposal\,\cite{Mueller} the atoms in an atom interferometer
are considered as ``clocks'' ticking at the Compton frequency
$\omega_\text{C}=m c^2/\hbar$ associated with their rest mass, and
propagating along two ``classical'' arms of the
interferometer. However we dispute this
interpretation\,\cite{WolfNat,WolfCQG}: (i) An atom is not a ``Compton
clock'', since it does not deliver a physical signal at the Compton
frequency\,\cite{WolfCQG,Samuel}.  (ii) In an atom interferometer, we
are using an interference between two possible paths followed by the
\textit{same} atom, which is described quantum mechanically. Contrary
to clock experiments the motion of atoms is not monitored. It is
deduced from the theory by using the same evolution equations which
allow one to evaluate the phase shift. Thus the two classical paths in
the interferometer cannot be determined by a measurement. In an
interferometer, where a single atom can propagate along two different
paths, trying to measure the path which is followed by the atom
destroys the interference signal (wave-particle
complementarity). (iii) Using the theory in a consistent manner, the
contribution to the phase shift which depends on the mass of the atom
(and therefore on its Compton frequency), and includes a contribution
from the gravitational redshift, is in fact exactly
zero\,\cite{Storey,Borde08,Wolf}.

The C{\ae}sium (or some alkali) atoms are optically cooled and
launched in a vertical fountain geometry. They are prepared in a
hyperfine ground state $g$. A sequence of vertical laser pulses
resonant with a $g \rightarrow g'$ hyperfine transition is applied to
the atoms during their ballistic (i.e. free fall) flight. In the
actual experiments the atoms undergo a two-photon Raman transition
where the two Raman laser beams are counter-propagating. This results
in a recoil velocity of the atoms, with the effective wave vector $k$
transferred to the atoms being the sum of the wave vectors of the
counter-propagating lasers\,\cite{Borde89,Storey,Wolf}. A first pulse
at time $t=0$ splits the atoms into a coherent superposition of
hyperfine states $gg'$ with the photon recoil velocity yielding a
spatial separation of the two wave packets. A time interval $T$ later
the two wave packets are redirected toward each other by a second
laser pulse thereby exchanging the internal states $g$ and
$g'$. Finally a time interval $T'$ later the atomic beams recombine
and a third pulse is applied. After this pulse the interference
pattern in the ground and excited states is measured.

The calculation of the phase shift $\Delta \varphi$ of the atomic
interferometer in the presence of a gravitational field proceeds in
several steps\,\cite{Borde89,Kasevich,Storey}. The first contribution
to the phase shift comes from the free propagation of the atoms in the
two paths. Since atom interferometers are close to the classical
regime, a path integral approach is very appropriate as it reduces to
a calculation of integrals along classical paths for a Lagrangian
which is at most quadratic in position $z$ and velocity $\dot{z}$,
i.e. is of the general type\,\cite{Storey}
\begin{equation}\label{quadratic}
L\left[z,\dot{z}\right] =
a(t)\,\dot{z}^2+b(t)\,\dot{z}z+c(t)\,z^2+d(t)\,\dot{z}+e(t)\,z+f(t)\,,
\end{equation}
where $a(t)$, $b(t)$, $c(t)$, $d(t)$, $e(t)$ and $f(t)$ denote some
arbitrary functions of time $t$. The phase shift due to the free
propagation of the atoms is given by the \textit{classical} action
\begin{equation}\label{Scl}
S_\text{cl}(z_T,T;z_0,0) = \int_{0}^{T} \ud t
\,L\left[z_\text{cl}(t),\dot{z}_\text{cl}(t)\right]\,,
\end{equation}
where the integral extends over the classical path $z_\text{cl}(t)$
obeying the Lagrange equations, with boundary conditions
$z_\text{cl}(0)=z_0$ and $z_\text{cl}(T)=z_T$.  Thus the phase
difference due to the free propagation of the atoms in the
interferometer is equal to the difference of classical actions in the
two paths,
\begin{equation}\label{action_phase}
  \Delta \varphi_S \equiv \frac{\Delta S_\text{cl}}{\hbar} = \frac{1}{\hbar}\oint \ud t
  \,L\left[z_\text{cl},\dot{z}_\text{cl}\right] \,,
\end{equation}
where we use the notation $\oint\ud\tau$ to mean the difference of
integrals between the two paths of the interferometer, assumed to form
a close contour.

\bigskip\noindent
\textbf{Theorem.}\,\cite{Borde89,Kasevich,Storey,Wolf,AB03a,AB03b,Borde08,WolfCQG}
\textit{For any quadratic Lagrangian of the form \eqref{quadratic} the
  difference of classical actions in the interferometer, and therefore
  the phase shift due to the free propagation of the atoms, reduces to
  the contribution of the change of internal states $g$ and $g'$,
  thus}\,\footnote{Rigorously, in this equation the time interval
  should be a proper time interval.}
\begin{equation}\label{Dphiresult}
\Delta S_\text{cl} = E_{gg'}(T-T')\,,
\end{equation}
\textit{where the internal energy change is denoted by $E_{gg'}\equiv
  E_{g'}-E_{g}$. (In particular, when the interferometer is symmetric,
  which will be the case for a closed Mach-Zehnder geometry, and for
  $T=T'$, we get exactly $\Delta S_\text{cl}=0$.)}

\medskip One calculates the classical trajectories of the wave packets
in the two arms using the equations of motion of massive test bodies
deduced from the classical Lagrangian and the known boundary
conditions (position and momenta) of the wave packets. In the case of
the general quadratic Lagrangian \eqref{quadratic} the equations of
motion read
\begin{equation}\label{eom}
\frac{\ud}{\ud t}\Bigl[2a(t) \dot{z}\Bigr] = \bigl[2c(t) -
\dot{b}(t)\bigr] z + e(t) -\dot{d}(t)\,.
\end{equation}
Then, one calculates the difference in the classical action integrals
along the two paths. Denoting by $z_1(t)$ and $z_3(t)$ the classical
trajectories between the laser interactions in the upper path, and by
$z_2(t)$ and $z_4(t)$ the trajectories in the lower path, we have
\begin{equation}\label{DS}
  \Delta S_\text{cl} = \int_0^T \Bigl(L[z_1,\dot{z}_1]-L[z_2,\dot{z}_2]\Bigr)
  \ud t
  + \int_T^{T+T'} \Bigl(L[z_3,\dot{z}_3]-L[z_4,\dot{z}_4]\Bigr)\ud t
  + E_{gg'}(T-T')\,,
\end{equation}
where the integrals are carried out along the classical paths
calculated in the first step. We have taken into account the changes
in energy $E_{gg'}$ between the hyperfine ground states $g$ and $g'$
of the atoms in each path. These energies will cancel out from the two
paths provided that $T'$ is equal to $T$, which will be true for a
Lagrangian in which we neglect gravity gradients\,\cite{Wolf}.

We now show that the two action integrals in \eqref{DS} cancel each
other in the case of the quadratic Lagrangian \eqref{quadratic}. This
follows from the fact that the difference between the Lagrangians
$L[z_1(t),\dot{z}_1(t)]$ and $L[z_2(t),\dot{z}_2(t)]$, which are
evaluated at the same time $t$ but on two different trajectories
$z_1(t)$ and $z_2(t)$, is a total time-derivative when the Lagrangians
are ``on-shell'', i.e. when the two trajectories $z_1(t)$ and $z_2(t)$
satisfy the equations of motion \eqref{eom}. To prove this we consider
the difference of Lagrangians $L_1-L_2\equiv L[z_1,\dot{z}_1] -
L[z_2,\dot{z}_2]$ on the two paths, namely
\begin{equation}\label{L12}
  L_1-L_2 = a\,(\dot{z}^2_1-\dot{z}^2_2) + b\,(\dot{z}_1z_1-\dot{z}_2z_2) 
  + c\,(z^2_1-z^2_2) + d\,(\dot{z}_1-\dot{z}_2) + e\,(z_1-z_2)\,.
\end{equation}
We re-express the first contribution $a(\dot{z}^2_1-\dot{z}^2_2)$
thanks to an integration by parts as $a(\dot{z}^2_1-\dot{z}^2_2) =
\frac{\ud}{\ud
  t}[a(z_1-z_2)(\dot{z}_1+\dot{z}_2)]-(z_1-z_2)\frac{\ud}{\ud
  t}[a(\dot{z}_1+\dot{z}_2)]$. The second term is then simplified by
means of the sum of the equations of motion \eqref{eom} written for
$z=z_1$ and $z=z_2$. In addition we also integrate by parts the second
and fourth contributions in \eqref{L12} as
$b(\dot{z}_1z_1-\dot{z}_2z_2)=\frac{\ud}{\ud
  t}[\frac{1}{2}b(z^2_1-z^2_2)]-\frac{1}{2}\dot{b}(z^2_1-z^2_2)$ and
$d(\dot{z}_1-\dot{z}_2)=\frac{\ud}{\ud
  t}[d(z_1-z_2)]-\dot{d}(z_1-z_2)$. Summing up the results we obtain
\begin{equation}\label{totaltimeder}
  L_1-L_2 = \frac{\ud}{\ud
    t}\biggl[(z_1-z_2)\biggl(a\,(\dot{z}_1+\dot{z}_2)+\frac{1}{2}b\,(z_1+z_2)
  +d\biggr)\biggr]\,.
\end{equation}
Since the difference of Lagrangians is a total time derivative the
difference of action functionals in \eqref{DS} can be immediately
integrated. Using the continuity conditions at the interaction points
with the lasers \eqref{cont}, which are
\begin{subequations}\label{cont}{\begin{align}
& z_1(0) = z_2(0)\,,\\
& z_1(T) = z_3(T)\,,\\
& z_2(T) = z_4(T)\,,\\
& z_3(T+T') = z_4(T+T')\,,
\end{align}}\end{subequations}
and are appropriate to a closed-path interferometer which closes up at
time $T+T'$, we obtain
\begin{equation}\label{DSres}
  \Delta S_\text{cl} =
  a(T)\bigl[z_1(T)-z_2(T)\bigr]\Bigl[\dot{z}_1(T)
  +\dot{z}_2(T)-\dot{z}_3(T)-\dot{z}_4(T)\Bigr] + E_{gg'}(T-T')\,.
\end{equation}
Next we apply the boundary conditions in velocities which are
determined by the recoils induced from the interactions with the
lasers. We see that once we have imposed the closure of the two paths
of the interferometer, only the recoils due to the second pulse at the
intermediate time $T$ are needed for this calculation. These are given
by
\begin{subequations}\label{deltav}{\begin{align}
\dot{z}_1(T)-\dot{z}_3(T)&=+\frac{\hbar k}{m}\,,\\
\dot{z}_2(T)-\dot{z}_4(T)&=-\frac{\hbar k}{m}\,,
\end{align}}\end{subequations}
where $k$ is the effective wave vector transferred by the lasers to
the atoms. This readily shows that the first term in \eqref{DSres} is
zero for any quadratic Lagrangian hence $\Delta S_\text{cl} =
E_{gg'}(T-T')$. $\blacksquare$

Finally, one calculates the contribution to the phase shift due to the
light phases of the lasers. These are obtained using the paths
calculated previously and the equations of light propagation, with the
light acting as a ``ruler'' that measures the motion of the atoms.
The phase difference from light interactions $\Delta\varphi_\ell$ is a
sum of terms given by the phases $\phi$ of the laser light as seen by
the atom, i.e. $\phi(z,t) = k z - \omega t - \phi_0$ where $k$,
$\omega$ and $\phi_0$ are the wave vector, frequency and initial phase
of the laser in the frame of the laboratory, and evaluated at all 
the interaction points with the lasers\,\cite{Borde89,Wolf}. Finally the
total phase shift measured in the atom interferometer is
\begin{equation}\label{DeltaAll}
  \Delta\varphi = \omega_{gg'}(T-T') + \Delta \varphi_\ell\,,
\end{equation}
and depends only on the internal states $g$ and $g'$ through
$\omega_{gg'}=E_{gg'}/\hbar$, and the light phases which measure the
free fall trajectories of the atoms. At the Newtonian approximation in
a uniform gravitational field $g$, the interferometer is symmetric,
$T'=T$, and one finds\,\cite{Borde89}
\begin{equation}\label{DeltaRG}
  \Delta\varphi = \Delta \varphi_\ell = k\,g\,T^2\,.
\end{equation}
This clearly shows that the atom interferometer is a
\textit{gravimeter} (or accelerometer): It measures the acceleration
$g$ of atoms with respect to the experimental platform which holds the
optical and laser elements. With $k$ and $T$ known from auxiliary
measurements, one deduces the component of $g$ along the direction of
$k$. If the whole instrument was put into a freely falling laboratory,
the measured signal $\Delta\varphi$ would vanish.

The result for the final phase shift \eqref{DeltaAll} or
\eqref{DeltaRG} is valid whenever the result \eqref{Dphiresult} holds,
i.e. in all theories of gravity defined by a single (quadratic)
Lagrangian and consistent with the principle of least action. In such
theories the Feynman path integral formulation of quantum mechanics
remains valid, and a coherent analysis of atom interferometry
experiments is possible. Most alternative theories commonly considered
belong to this class which encompasses a large number of models and
frameworks\,\cite{Will}.  It includes for example most non-metric
theories, some models motivated by string theory\,\cite{DamourP} and
brane scenarios, some general parameterized frameworks such as the
energy conservation formalism\,\cite{Nordtvedt,Haugan}, the
$\text{TH}\varepsilon\mu$ formalism\,\cite{LightLee}, and the Lorentz
violating standard model extension (SME)\,\cite{Koste1,Koste2}. In all
such theories the action-phase shift of the atom interferometer is
zero (in particular the Compton frequency of the atom is
irrelevant). Because there is no way to disentangle the gravitational
redshift from the Doppler shift, we conclude that the recent
proposal\,\cite{Mueller} is invalidated.


\vspace{1cm}
\section*{References}

\begin{thebibliography}{99}

\bibitem{Will} C.~M. Will, \textit{Theory and experiment in
    gravitational physics} (Cambridge University Press, 1993).

\bibitem{Dicke} P.~Roll, R.~Krotkov, and R.~Dicke, Ann. Phys. (N.Y.)
  \textbf{26}, 442 (1964).

\bibitem{Braginsky} V.~Braginsky and V.~Panov, Sov. Phys. JETP
  \textbf{34}, 463 (1972).

\bibitem{Schlamminger} S.~Schlamminger, K.-Y. Choi, T.~Wagner,
  J.~Gundlach, and E.~Adelberger, Phys. Rev. Lett. \textbf{100},
  041101 (2008).

\bibitem{Williams} J.~Williams, S.~Turyshev, and D.~Boggs,
  Phys. Rev. Lett. \textbf{93}, 261101 (2004).

\bibitem{Peters} A.~Peters, K.~Chung, and S.~Chu, Nature \textbf{400},
  849 (1999).

\bibitem{refsyrte} S.~Merlet, Q.~Bodart, N.~Malossi, A.~Landragin, and
  F.~Pereira Dos~Santos, Metrologia \textbf{47}, L9 (2010).

\bibitem{Vessot} R.~Vessot and M.~Levine, Gen. Rel. and
  Grav. \textbf{10}, 181 (1979).

\bibitem{Cacciapuoti} L.~Cacciapuoti and C.~Salomon,
  Eur. Phys. J. Spec. Top. \textbf{127}, 57 (2009).

\bibitem{WolfNat} P.~Wolf, L.~Blanchet, Ch.J.~Bord{\'e}, S.~Reynaud,
  C.~Salomon, and C.~Cohen-Tannoudji, Nature \textbf{467}, E1 (2010),
  arXiv:1009.0602 [gr-qc].

\bibitem{WolfCQG} P.~Wolf, L.~Blanchet, Ch.J.~Bord{\'e}, S.~Reynaud,
  C.~Salomon, and C.~Cohen-Tannoudji, Class. Quant. Grav. \textbf{28},
  145017 (2011), arXiv:1009.2485 [gr-qc].

\bibitem{Mueller} H.~M{\"u}ller, A.~Peters, and S.~Chu, Nature
  \textbf{463}, 926 (2010).

\bibitem{Samuel} S.~Sinha and J.~Samuel,
  Class. Quant. Grav. \textbf{28}, 145018 (2011), arXiv:1102.2587
  [gr-qc].

\bibitem{Giulini} D.~Giulini (2011), arXiv:1105.0749 [gr-qc].

\bibitem{Pound} R.~Pound and G.~Rebka, Phys. Rev. Lett. \textbf{4},
  337 (1960).

\bibitem{Borde89} Ch.J.~Bord\'e, Phys. Lett. A \textbf{140}, 10
  (1989).

\bibitem{Storey} P.~Storey and C.~Cohen-Tannoudji, J. Phys. II France
  \textbf{4}, 1999 (1994).

\bibitem{Wolf} P.~Wolf and P.~Tourrenc, Phys. Lett. A \textbf{251},
  241 (1999).

\bibitem{AB03a} Ch. Antoine and Ch.J.~Bord\'e,
  Phys. Lett. A \textbf{306}, 277 (2003).

\bibitem{AB03b} Ch. Antoine and Ch.J.~Bord\'e,
  J. Opt. B \textbf{5}, S199 (2003).

\bibitem{Borde08} Ch.J.~Bord\'e,
  Eur. Phys. J. Spec. Top. \textbf{163}, 315 (2008).

\bibitem{Kasevich} M.~Kasevich and S.~Chu,
  Phys. Rev. Lett. \textbf{67}, 181 (1991).

\bibitem{DamourP} T.~Damour and A.~Polyakov, Nucl. Phys. B
  \textbf{423}, 532 (1994).

\bibitem{Nordtvedt} K.~Nordtvedt, Phys. Rev. D \textbf{11}, 245
  (1975).

\bibitem{Haugan} M.~Haugan, Ann. Phys. (N.Y.) \textbf{118}, 156
  (1979).

\bibitem{LightLee} A.~Lightman and D.~Lee, Phys. Rev. D \textbf{8},
  364 (1973).

\bibitem{Koste1} Q.~Bailey and V.~Kostelecky, Phys. Rev. D
  \textbf{74}, 045001 (2006).

\bibitem{Koste2} V.~Kostelecky and J.~Tasson (2010), arXiv:1006.4106.

\end{thebibliography}
\end{document}